\documentclass[12pt,fleqn]{article}
\usepackage{graphics,epsfig}

\newcommand{\blackslug}{\rule{7pt}{7pt}}
\newcommand{\ignore}[1]{}
\newcommand{\beq}[1]{\begin{equation}\label{#1}}
\newcommand{\eeq}{\end{equation}}
\newcommand{\beqa}[1]{\begin{equation}\label{#1}\begin{eqalign}}
\newcommand{\eeqa}{\end{eqalign}\end{equation}}
\newcommand{\bsubeq}[1]{\begin{subequations}\label{#1}\begin{eqalignno}}
\newcommand{\esubeq}{\end{eqalignno}\end{subequations}}
\newenvironment{proof}{{\bf Proof.}\/}{\hspace*{\fill} \qed \vskip 0.1in}

\def \X   {{ X}}
\def \tX   {{\tilde X}}
\def \x   {{\tilde x}}
\def \Y  {{ Y}}

\def \F  {{\cal F}}
\def \L  {{\cal L}}
\def \b1  {{\beta^{-1}}}
\def \l  {{\lambda(x)}}
\def\ket{\rangle}
\def\bra{\langle}
\newcommand{\comment}[1]{}

\newcommand{\real}{\ifmmode {\rm R} \else ${\rm R}$ \fi}
\newcommand{\nat}{\ifmmode {\rm N} \else ${\rm N}$  \fi}
\newcommand{\tot}{\ifmmode {\cal T} \else ${\cal T}$ \fi}
\newcommand{\sigstar}{\ifmmode \Sigma^{\ast} \else $\Sigma^{\ast}$ \fi}

\newcommand{\inn}{\ifmmode \in \else $\in$ \fi}
\renewcommand{\phi}{\ifmmode \varphi \else $\varphi$ \fi}
\renewcommand{\le}{\ifmmode \leq \else $\leq$ \fi}
\renewcommand{\ge}{\ifmmode \geq \else $\geq$ \fi}
\renewcommand{\ne}{\ifmmode \neq \else $\neq$ \fi}
\newcommand{\lt}{\ifmmode < \else $<$ \fi}
\newcommand{\gt}{\ifmmode > \else $>$ \fi}
\newcommand{\eq}{\ifmmode = \else $=$ \fi}
\newcommand{\half}{\ifmmode \frac{1}{2} \else $\frac{1}{2}$ \fi}
\newcommand{\oneovern}{\ifmmode \frac{1}{n} \else $\frac{1}{n}$ \fi}

\newcommand{\ra}{\ifmmode \rightarrow \else $\rightarrow$ \fi}
\newcommand{\imp}{\ifmmode \Rightarrow \else $\Rightarrow$ \fi}
\newcommand{\implies}{\ifmmode \Rightarrow \else $\Rightarrow$ \fi}
\newcommand{\qed}{\hfill{\setlength{\fboxsep}{0pt}
\framebox[7pt]{\rule{0pt}{7pt}}}}

\renewcommand{\notin}{\ifmmode \not\in \else $\not\in$ \fi}
\newtheorem{theorem}{Theorem}
\newtheorem{lemma}[theorem]{Lemma}

\newlength{\thislabel}

\begin{document}
\baselineskip = 16 pt
\leftline{\Large\bf The information bottleneck method}
\bigskip\bigskip

\leftline{\large Naftali Tishby,$^{1,2}$ Fernando C. Pereira,$^3$ and
William Bialek$^1$}
\bigskip

\leftline{$^1$NEC Research Institute, 4 Independence Way} 
\leftline{Princeton, New Jersey 08540}
\leftline{$^2$Institute for Computer Science, and}
\leftline{Center for Neural Computation}
\leftline{Hebrew University}
\leftline{Jerusalem 91904, Israel}
\leftline{$^3$AT\&T Shannon Laboratory}
\leftline{180 Park Avenue}
\leftline{Florham Park, New Jersey 07932}
\bigskip

\leftline{30 September 1999}

\bigskip\bigskip\hrule\bigskip\bigskip

We define the relevant information in a signal $x\in X$ as being the
information that this signal provides about another signal $y\in \Y$.
Examples include the information that face images provide about the
names of the people portrayed, or the information that speech sounds
provide about the words spoken.  Understanding the signal $x$ requires
more than just predicting $y$, it also requires specifying which
features of $\X$ play a role in the prediction.  We formalize this
problem as that of finding a short code for $\X$ that preserves the
maximum information about $\Y$.  That is, we squeeze the information
that $\X$ provides about $\Y$ through a `bottleneck' formed by a
limited set of codewords $\tX$.  This constrained optimization problem
can be seen as a generalization of rate distortion theory in which the
distortion measure $d(x,\x)$ emerges from the joint statistics of $\X$
and $\Y$.  This approach yields an exact set of self consistent
equations for the coding rules $X \rightarrow \tX$ and $\tX
\rightarrow \Y$.  Solutions to these equations can be found by a
convergent re--estimation method that generalizes the Blahut--Arimoto
algorithm.  Our variational principle provides a surprisingly rich
framework for discussing a variety of problems in signal processing
and learning, as will be described in detail elsewhere.

\vfill\newpage

\section{Introduction}

A fundamental problem in formalizing
our intuitive ideas about information is to provide a quantitative
notion of ``meaningful'' or ``relevant''
information.   These issues were
intentionally left out of information theory in its original
formulation by Shannon, who focused attention on the problem of
transmitting
information rather than judging its value to the recipient.
Correspondingly, information theory has often been viewed
as being strictly a theory of
communication, and this view
has become so accepted that many people  consider statistical and
information
theoretic principles as almost irrelevant for the question of meaning.
In contrast, we argue here that
information theory, in particular lossy source compression,
provides a natural quantitative approach to the question of
``relevant information.'' Specifically, we formulate a variational
principle for the extraction or efficient representation of relevant
information.  In related work \cite{Bialek_Tishby99} we argue  that this
single information theoretic principle contains as special cases
a wide variety of problems, including prediction, filtering, and learning
in its various forms.

The problem of extracting a relevant summary of data, a compressed
description that captures only the relevant or meaningful
information,
is not well posed without a suitable definition of  relevance.
A typical example is that of speech compression. One can
consider lossless compression, but in any compression beyond the entropy
of speech some components of the signal cannot be reconstructed.
On the other hand, a transcript of the spoken words has much lower
entropy
(by orders of magnitude) than the acoustic waveform, which means that
it is possible to compress (much) further without losing any information
about the words and their meaning.

The standard analysis of lossy source compression is
``rate distortion theory,'' which characterizes the tradeoff between the
rate, or signal representation size, and the average distortion of the
reconstructed signal. Rate distortion theory determines the level
of inevitable expected distortion, $D$, given the desired information
rate, $R$, in terms of the {\em rate distortion function} $R(D)$.
The main problem with rate distortion theory is in the need to specify
the distortion function first, which in turn
determines the relevant features of the signal. Those features,
however, are often not explicitly known and an arbitrary choice of the
distortion function is in fact an arbitrary feature selection.

In the speech example, we have at best very partial
knowledge of what precise components of the signal are perceived by our
(neural) speech recognition system.
Those relevant components
depend not only on the complex structure of the auditory nervous
system, but also on the sounds and utterances to which we are exposed
during our early life. It therefore is virtually impossible to come
up with the ``correct'' distortion function for acoustic signals. The
same type of difficulty exists in many similar problems, such as
natural language processing, bioinformatics
(for example, what features of
protein sequences determine their structure) or neural coding
(what information is encoded by spike trains and how). This is
the fundamental problem of feature selection in pattern recognition.
Rate distortion theory does not provide a full answer to this
problem since the choice of the distortion function, which determines
the relevant features, is not part of the theory.
It is, however, a step in the right direction.

A possible solution comes from the fact that in many interesting cases
we have access to an additional variable that determines what is
relevant.
In the speech case it might be the transcription of the signal,
if we are interested in the speech recognition problem, or it might be the
speaker's identity if speaker identification is our goal.
For natural language processing, it might be the part of speech labels
for words in grammar checking, but the dictionary senses of ambiguous
words in information retrieval. Similarly, for the protein folding
problem we have a joint database of sequences and
three dimensional structures,  and for
neural coding a simultaneous recording of sensory stimuli and neural
responses defines implicitly the relevant variables in each domain.
All of these problems have
the same formal underlying structure: extract the information from one
variable that is relevant for the prediction of another one.
The choice of additional variable determines the relevant
components or features of the signal in each case.

In this short paper we formalize this intuitive idea using an information
theoretic approach which extends elements of rate distortion theory.
We derive self consistent equations
and an iterative algorithm for finding representations of the signal
that capture its relevant structure, and prove convergence of this
algorithm.

\section{Relevant quantization}

Let $\X$ denote the signal (message) space with a fixed probability
measure $p(x)$, and let $\tX$ denote its quantized codebook or
compressed representation.  For ease of exposition we assume here that
both of these sets are finite, that is, a continuous space should
first be quantized.

For each value $x \in \X$ we seek a
possibly stochastic mapping to a representative, or codeword in a
codebook, $\x \in \tX$,
characterized by a conditional p.d.f. $p(\x|x)$. The mapping $p(\x|x)$
induces a soft partitioning of $\X$ in which each block is associated
with one of the codebook elements $\x \in \tX$, with probability given
by
\begin{equation}
\label{r1}
p(\x) =\sum_x p(x) p(\x|x) ~.
\end{equation}
The average
volume of the elements of $\X$ that are mapped to the same codeword
is $2^{H(\X|\tX)}$, where
\begin{equation}
H(\X|\tX) = -\sum_{x\in \X}p(x) \sum_{\x\in\tX} p(\x|x) \log p(\x|x)
\end{equation}
is the conditional entropy of $\X$ given $\tX$.

What determines the quality of a quantization? The first factor is of
course the rate, or the average number of bits per message
needed to specify an element in the codebook without confusion.
This number {\em per element of $\X$} is  bounded from below by the
mutual information
\begin{equation}
I(\X;\tX) =\sum_{x\in \X}\sum_{\x\in\tX} p(x,\x) \log
\left[\frac{p(\x|x)}{p(\x)}\right] ,
\end{equation}
since the average cardinality of the partitioning of $\X$
is given by the ratio of the volume of $\X$ to that of the mean
partition,
$2^{H(\X)}/2^{H(\X|\tX)}=2^{I(\X;\tX)}$, via the standard asymptotic
arguments. Notice that this quantity is different from the entropy
of the codebook, $H(\tX)$, and this entropy  normally is not what we want
to minimize.

However, information rate alone is not enough to characterize good
quantization since the rate can always be reduced by throwing away
details of the original signal $x$.
We need therefore some additional constraints.

\subsection{Relevance through distortion:\\ Rate distortion theory}

In rate distortion theory such a constraint is provided through a
distortion function,
$d:\X \times \tX \rightarrow {\cal R}^+$, which is presumed to be
small for good representations. Thus the distortion function
specifies implicitly what are the most relevant aspects of values in
$\X$.

The  partitioning of $\X$ induced by the mapping $p(\x|x)$ 
has an expected distortion
\begin{equation}
\bra d(x,\x) \ket_{p(x,\x)} = \sum_{x\in \X}\sum_{\x\in\tX} p(x,\x)
d(x,\x) ~.
\end{equation}
There is a monotonic tradeoff between the rate of the quantization
and the expected distortion: the larger the rate, the smaller is the
achievable distortion.

The celebrated rate distortion theorem of Shannon and Kolmogorov (see,
for
example Ref. \cite{Cover-Thomas-91}) characterizes this tradeoff through
the rate
distortion function, $R(D)$, defined as the minimal achievable rate
under a given constraint on the expected distortion:
\begin{equation}
R(D) \equiv \min_{\{p(\x|x) : \bra d(x,\x) \ket \le D \}}
I(\X;\tX) ~.
\end{equation}
Finding the rate distortion function is a
variational problem that can be solved by introducing a
Lagrange multiplier, $\beta$, for the constrained expected distortion.
One then needs to minimize the functional
\begin{equation}
\label{RD1}
\F[p(\x|x)] = I(\X;\tX) + \beta \bra d(x,\x) \ket_{p(x,\x)}
\end{equation}
over all normalized distributions $p(\x|x)$.
This variational formulation has the following well known consequences:
\begin{theorem}
\label{var1}
The solution of the variational problem,
\begin{equation}
\label{va1}
\frac{\delta \F}{\delta p(\x|x)}=0,
\end{equation}
for normalized distributions $p(\x|x)$, is given by the exponential form
\begin{equation}
\label{va2}
p(\x | x)
= {{p(\x)}\over {Z(x,\beta)}} \exp\left[-\beta d(x,\x) \right] ~,
\end{equation}
where $Z(x,\beta)$ is a normalization (partition) function.
Moreover, the Lagrange multiplier $\beta$, determined by the
value of the expected distortion, $D$, is positive and
satisfies
\begin{equation}
\label{va3}
\frac{\delta R}{\delta D}=-\beta ~.
\end{equation}
\end{theorem}
\begin{proof}
Taking the derivative w.r.t. $p(\x|x)$, for given $x$ and $\x$, one
obtains
\begin{eqnarray}
\frac{\delta \F}{\delta p(\x|x)}
  & = & p(x)  \Bigg[ \log \frac{p(\x|x)}{p(\x)} +1  
\nonumber\\
&&\,\,\,\,\,\,\,\,\,\,\,\,\,\,\,\,\,\,\,\,
 -
{{1}\over{p(\x)}} \sum_{x'} p(x')p(\x|x')+\beta d(x,\x)
   +\frac{\l}{p(x)}
\Bigg] ,
\label{va4}
\end{eqnarray}
since the marginal distribution satisfies $p(\x)=\sum_{x'}
p(x')p(\x|x')$. $\l$ are the normalization Lagrange
multipliers for each $x$. Setting the derivatives to zero and writing
$\log Z(x,\beta) =\lambda(x)/p(x)$, we obtain Eq. (\ref{va2}).
When varying the normalized $p(\x|x)$, the variations $\delta I(\X;\tX)$
and
$\delta  \bra d(x,\x) \ket_{p(x,\x)}$ are linked through
\begin{equation}
\label{va5}
\delta \F = \delta  I(\X;\tX) + \beta \delta \bra d(x,\x)
\ket_{p(x,\x)}=0,
\end{equation}
from which Eq. (\ref{va3}) follows. The positivity of $\beta$ is then
a consequence of the concavity of the rate distortion function (see, for
example, Chapter 13 of Ref.
\cite{Cover-Thomas-91}).
\end{proof}

\subsection{The Blahut--Arimoto algorithm}

An important practical consequence of the above variational
formulation is that it provides a converging iterative algorithm for
self consistent determination of the distributions $p(\x|x)$ and $p(\x)$.

Equations (\ref{va2}) and (\ref{r1}) must be satisfied
simultaneously for consistent probability assignment. A natural
approach to solve these equations is to alternately iterate between
them until reaching convergence.
The following lemma, due to Csisz\'{a}r and Tusn\'{a}dy \cite{CT84},
assures global convergence in this case.
\begin{lemma}
\label{CT1}
Let $p(x,y) = p(x)p(y|x)$ be a given joint distribution.
Then the distribution
$q(y)$ that minimizes the relative entropy or Kullback--Leibler
divergence,
$D_{KL}[p(x,y)|p(x)q(y)]$, is the marginal distribution 
$$p(y)=\sum_x
p(x)p(y|x).$$
Namely,
$$
I(X;Y)=D_{KL}[p(x,y)|p(x)p(y)] = \min_{q(y)} D_{KL}[p(x,y)|p(x)q(y)] ~.
$$
Equivalently, the distribution $q(y)$ which minimizes the expected
relative entropy,
$$\sum_x p(x) D_{KL}[p(y|x)|q(y)],$$
is also the marginal
distribution $p(y)=\sum_x p(x)p(y|x)$.
\end{lemma}
The proof follows directly from the non--negativity of the relative
entropy.

This lemma guarantees the marginal condition
Eq. (\ref{r1}) through the same variational principle that leads to
Eq. (\ref{va2}):
\begin{theorem}
Equations (\ref{r1}) and (\ref{va2}) are satisfied
simultaneously at the minimum of the functional,
\begin{equation}
\F = -\bra \log Z(x,\beta)\ket_{p(x)} = I(\X;\tX) +
\beta \bra d(x,\x) \ket_{p(x,\x)} ~,
\end{equation}
where the minimization is done {\em independently} over the
convex sets of the normalized distributions, $\{p(\x)\}$ and
$\{p(\x|x)\}$,
\[
\min_{p(\x)} \min_{p(\x|x)} \F\left[ p(\x);p(\x|x) \right] ~.
\]
These independent conditions  correspond precisely
to alternating iterations of Eq. (\ref{r1}) and Eq. (\ref{va2}).
Denoting by $t$ the iteration step,
\begin{equation}
\label{BAalg}
\left\{
\begin{array}{l}
p_{t+1}(\x) = \sum_x p(x) p_t(\x|x)\\
p_t(\x|x) = \frac{p_t(\x)}{Z_t(x,\beta)} \exp(-\beta d(x,\x))
\end{array}
\right .
\end{equation}
where the normalization function $Z_t(x,\beta )$ is evaluated for
every $t$ in Eq. (\ref{BAalg}).
Furthermore, these iterations converge to a unique minimum of
$\F$ in the convex sets of the two distributions.
\end{theorem}
For the proof, see references \cite{Cover-Thomas-91,Blahut72}.
This alternating iteration is the well known Blauht-Arimoto
(BA) algorithm for calculation of the rate distortion function.

It is important to notice that the BA algorithm deals only with the
optimal partitioning of the set $\X$ given the set of representatives
$\tX$, and not with an optimal choice of the representation $\tX$. In
practice, for finite data, it is also important to find the optimal
representatives which minimize the expected distortion, {\em given}
the partitioning. This joint optimization is similar to the EM
procedure in statistical estimation and does not in general have a
unique solution.

\section{Relevance through another variable:\\ The Information Bottleneck}

Since the ``right'' distortion measure is rarely available, the problem
of relevant quantization has to be addressed directly, by preserving
the {\em relevant information} about another variable. The relevance
variable, denoted here by $\Y$, must not be independent from the
original signal $\X$, namely they have positive mutual information
$I(X;Y)$. It is assumed here that we have access to the joint
distribution $p(x,y)$, which is part of the setup of the problem,
similarly to $p(x)$ in the rate distortion case.\footnote{The problem of
actually obtaining a good enough sample of this distribution is an
interesting issue in learning theory, but is beyond the scope of this
paper.  For a start on this problem see Ref. \cite{Bialek_Tishby99}.}

\subsection{A new variational principle}

As before, we would like our relevant quantization $\tX$ to compress
$X$ as much as possible.
In contrast to the rate distortion problem, however, we now
want this quantization to capture as much of the
information about $Y$ as possible.
The amount of information about $\Y$ in $\tX$ is given by
\begin{equation}
I({\tilde X}; Y) = \sum_y \sum_{\x} p(y,\x)
\log\frac{p(y,\x)}{p(y)p(\x)}
\le I(\X;\Y).
\end{equation}
Obviously lossy compression cannot convey more information than the
original data.
As with rate and distortion, there is a tradeoff
between compressing the representation and preserving
meaningful information, and there is no single right solution for the
tradeoff.
The assignment we are looking for is the one that keeps a fixed amount of
meaningful information about the relevant signal $\Y$  while
minimizing the number of bits from the original signal $\X$
(maximizing the compression). \footnote{It is completely equivalent
to maximize the meaningful information for a fixed compression
of the original variable.}   In effect we pass the information that $X$
provides about $Y$ through a ``bottleneck'' formed by the compact
summaries in $\tX$.

We can find the optimal assignment by minimizing the functional
\begin{equation}
\L[p(\x|x)] = I({\tilde X}; \X) - \beta I({\tilde X} ; \Y) ,
\label{varprin}
\end{equation}
where $\beta$ is the Lagrange multiplier attached to the constrained
meaningful information, while maintaining the normalization of the
mapping $p(\x|x)$ for every $x$.
At $\beta=0$  our quantization is the most sketchy
possible---everything is assigned to a single point---while
as $\beta \rightarrow\infty$ we are pushed
toward arbitrarily detailed quantization. By varying the (only)
parameter $\beta$ one can explore the tradeoff between the preserved
meaningful information and compression at various resolutions.
As we show elsewhere
\cite{Bialek_Tishby99,SloTish99},
for interesting
special cases (where there exist sufficient statistics) it is
possible to preserve almost all the meaningful information at finite
$\beta$ with a significant compression of the original data.

\subsection{Self-consistent equations}
Unlike the case of rate distortion theory, here the
constraint on the meaningful information is {\em nonlinear} in the
desired mapping $p(\x|x)$ and this is a much harder variational problem.
Perhaps surprisingly, this general problem of extracting the
meaningful information---minimizing the functional
$\L[p(\x|x)]$ in
Eq. (\ref{varprin})---can be given an exact formal solution.
\begin{theorem}
\label{Self-consistent}
The optimal assignment, that minimizes Eq. (\ref{varprin}),
satisfies the equation
\begin{equation}
p({\tilde x} | x)
= {{p({\tilde x})}\over Z(x,\beta)}
\exp\left[ - \beta \sum_y p(y|x) \log {{p(y|x)}\over{p(y|\x)}}\right]
,
\label{selfcon1}
\end{equation}
where the distribution $p(y|\x)$ in the exponent is given via
Bayes' rule and the Markov chain condition $\tX\leftarrow\X \leftarrow
\Y$, as,
\begin{equation}
p(y|{\tilde x}) = {1\over {p(\x)}}\sum_x p(y|x) p(\x|x)p(x) .
\label{selfcon2}
\end{equation}
\end{theorem}
This solution has a number of interesting features, but we must
emphasize that it is a {\em formal} solution since $p(y|{\tilde x})$ in
the
exponential is defined implicitly in terms of the
assignment mapping $p({\tilde x}|x)$. Just as before, the marginal
distribution $p(\x)$ must satisfy the marginal condition
Eq. (\ref{r1}) for consistency.

\begin{proof}
\label{Derivation_of_the_self-consistent_equations}
First we note that the conditional distribution of $y$ on $\x$
\beq{E1}
p(y|\x)= \sum_{x \in \X} p(y|x) p(x|\x) ~,
\eeq
follows from the Markov chain condition $\Y\leftarrow\X \leftarrow
\tX$.\footnote{It is important to notice that this not a
modeling assumption and the quantization $\tX$ is {\em not} used as
a hidden variable in a model of the data. In the latter, the Markov
condition would have been different: $\Y \leftarrow\tX \leftarrow \X$.}
The only variational variables in this scheme are the conditional
distributions, $p(\x|x)$, since other unknown distributions are
determined from it through Bayes' rule and consistency. Thus we have
\beq{E2}
p(\x) = \sum_x  p(\x|x) p(x)~,
\eeq
and
\beq{E3}
p(\x|y) = \sum_x  p(\x|x)p(x|y) ~.
\eeq
The above equations imply the following derivatives w.r.t. $p(\x|x)$,
\beq{E4}
\frac{\delta p(\x)}{\delta p(\x|x)} = p(x)
\eeq
and
\beq{E5}
\frac{\delta p(\x|y)}{\delta p(\x|x)} = p(x|y) ~.
\eeq
Introducing Lagrange multipliers, $\beta$ for the information
constraint and $\lambda(x)$ for the normalization of the conditional
distributions $p(\x|x)$ at each $x$, the Lagrangian,
Eq. (\ref{varprin}), becomes
\begin{eqnarray}
\label{E6}
\L &=& I(\X,\tX) - \beta I(\tX,\Y) - \sum_{x,\x} \lambda(x)
p(\x|x) \\ 
    &=&   \sum_{x,\x} p(\x|x)p(x)\log\left[\frac{p(\x|x)}{p(\x)}\right]
-\beta \sum_{\x,y} p(\x,y) \log \left[\frac{p(\x|y)}{p(\x)}\right]
\nonumber\\
&&\,\,\,\,\,\,\,\,\,\,
\,\,\,\,\,\,\,\,\,\,
- \sum_{x,\x} \l p(\x|x) ~.
\end{eqnarray}
Taking derivatives with respect to $p(\x|x)$
for given $x$ and $\x$,  one obtains
\begin{eqnarray}
\label{E7}
\frac{\delta \L}{\delta p(\x|x)}
&=& p(x)\left[1 + \log {p(\x|x)} \right]
-\frac{ \delta p(\x) }{ \delta p(\x|x) }
\left[1 + \log {p(\x)} \right]  
\nonumber\\
&&\,\,\,\,\,\,\,\,\,\,   -  \beta \sum_y \frac{\delta
p(\x|y)}{\delta p(\x|x)} {p(y)}
\left[ 1 + \log {p(\x|y)} \right]
\nonumber\\
&&\,\,\,\,\,\,\,\,\,
  - \beta\frac{\delta p(\x)}{\delta p(\x|x)}
\left[ 1 + \log {p(\x)} \right]  -\l ~.
\end{eqnarray}
Substituting the derivatives from Eq's. (\ref{E4}) and (\ref{E5})  and
rearranging, 
\begin{equation}
\label{E8}
\nonumber
\frac{\delta \L}{\delta p(\x|x)}
  = p(x) \left\{ \log \left[\frac{p(\x|x)}{p(\x)} \right]
   -\beta \sum_y p(y|x) \log \left[\frac{p(y|\x)}{p(y)}\right]
-\frac{\l}{p(x)}
\right\}  ~.
\end{equation}
Notice that $\sum_y p(y|x)\log\frac{p(y|x)}{p(y)} = I(x,\Y)$ is a
function of
$x$ only (independent of $\x$) and thus can be absorbed into the
multiplier $\l$. Introducing
$$
\tilde{\lambda}(x)= \frac{\l}{p(x)} -
\beta \sum_y p(y|x)\log\left[\frac{p(y|x)}{p(y)} \right] ~,
$$
we finally obtain the variational condition:
\begin{eqnarray}
\label{E9}
\frac{\delta \L}{\delta p(\x|x)} =
p(x) \left[\log \frac{p(\x|x)}{p(\x)}+
\beta \sum_{y} p(y|x)\log\frac{p(y|x)}{p(y|\x)}
-\tilde{\lambda}(x) \right] =0 ~,
\end{eqnarray}
which is equivalent to equation (\ref{selfcon1}) for $p(\x|x)$,
\begin{eqnarray}
\label{E10}
p(\x|x) =
\frac{p(\x)}{Z(x,\beta)} \exp \left( -\beta D_{KL} \left[p(y|x) |
p(y|\x) \right] \right) ~,
\end{eqnarray}
with
$$
Z(x,\beta) =\exp[\beta \tilde{\lambda}(x)] = \sum_{\x} p(\x)
\exp \left( -\beta D_{KL} \left[p(y|x) | p(y|\x) \right] \right) ~,
$$
the normalization (partition) function.
\end{proof}
{\bf Comments:}
\begin{enumerate}
\item
The Kullback--Leibler divergence, $ D_{KL}[p(y|x)|p(y|\x)]$, {\em emerged}
as the relevant ``effective distortion measure'' from our
variational principle but is not assumed otherwise anywhere! It is
therefore natural to consider it as the ``correct'' distortion
$d(x,\x)=D_{KL}[p(y|x)|p(y|\x)]$ for quantization in the
information bottleneck setting.
\item
Equation (\ref{E10}), together with equations (\ref{E1}) and (\ref{E2}),
determine self consistently the desired conditional distributions
$p(\x|x)$ and $p(\x)$.
The crucial quantization is here performed
through the conditional distributions $p(y|\x)$, and the
self consistent equations
include also the optimization over the representatives, in contrast to
rate distortion theory, where the selection of representatives is a
separate problem.
\end{enumerate}

\subsection{The information bottleneck iterative algorithm}

As for the BA algorithm, the self consistent equations
(\ref{selfcon1}) and (\ref{selfcon2})
suggest a natural method for finding the unknown distributions,
at every value of $\beta$. Indeed, these equations can be turned into
converging, alternating iterations among the three convex
distribution sets, $\{p(\x|x)\}$, $\{p(\x)\}$, and $\{p(y|\x)\}$, as
stated in the following theorem.

\begin{theorem}
The self consistent equations  (\ref{E1}), (\ref{E2}), and (\ref{E10}),
are satisfied simultaneously at the minima of the functional,
\begin{eqnarray}
\F\left[p(\x|x);p(\x);p(y|\x)\right]
&=& -\bra \log Z(x,\beta)\ket_{p(x)}\\
&=& I(\X;\tX) +
\beta \bra D_{KL}[p(y|x)|p(y|\x)] \ket_{p(x,\x)} ~,
\end{eqnarray}
where the minimization is done {\em independently} over the
convex sets of the normalized distributions, $\{p(\x)\}$ and
$\{p(\x|x)\}$ and $\{p(y|\x)\}$. Namely,
\[
\min_{p(y|\x)}
\min_{p(\x)} \min_{p(\x|x)} \F\left[p(\x|x); p(\x);p(y|\x) \right] ~.
\]
This minimization is performed by the converging alternating iterations.
Denoting by $t$ the iteration step,
\begin{equation}
\label{IBM}
\left\{
\begin{array}{l}
\label{iterateIBM1}
p_t(\x|x) = \frac{p_t(\x)}{Z_t(x,\beta)} \exp(-\beta d(x,\x))\\
p_{t+1}(\x) = \sum_x p(x) p_t(\x|x)\\
p_{t+1}(y| \x) = \sum_y p(y|x ) p_t(x|\x)
\end{array}
\right .
\end{equation}
and the normalization (partition function) $Z_t(\beta,\x)$ is evaluated
for
every $t$ in Eq. (\ref{iterateIBM1}).
\end{theorem}
\begin{proof}
For lack of space we can only outline the proof.
First we show that the equations
indeed are satisfied at the minima of the functional $\F$
(known for physicists as the ``free energy'').
This follows from lemma (\ref{CT1})  when applied to
$I(\X;\tX)$ with the convex sets of
$p(\x)$ and $p(\x|x)$, as for the BA algorithm.
Then the second part of the lemma is applied to
$\bra D_{KL}[p(y|x)|p(y|\x)] \ket_{p(x,\x)}$ which is an expected
relative entropy.  Equation (\ref{E10}) minimizes the expected relative
entropy w.r.t. to variations in the convex set of the 
normalized $\{p(y|\x)\}$. Denoting by
$d(x,\x)=D_{KL}[p(y|x)|p(y|\x)]$ and by $\lambda(\x)$
the normalization Lagrange multipliers, we obtain
\begin{eqnarray}
\label{delD}
\delta d(x , \x) &=& \delta \left( -\sum_y p(y|x ) \log p(y|\x)
+ \lambda(\x )( \sum_y p(y|\x) -1) \right)\\
&=& \sum_y \left( -\frac{p(y|x ) }{p(y|\x)} + \lambda(\x) \right)
\delta p(y|\x) ~.
\end{eqnarray}
The expected relative entropy becomes,
\beq{delD1}
  \sum_x \sum_y \left( -\frac{p(y|x ) p(x|\x )}{p(y|\x)}
+ \lambda(\x) \right) \delta p(y|\x) =0 ~,
\eeq
which gives Eq. (\ref{E10}), since $\delta p(y|\x)$ are independent
for each $\x$.
Equation (\ref{E10}) also have the interpretation of a
weighted average of the data conditional distributions
that contribute to the representative $\x$.

To prove the convergence of the iterations it is enough to verify that
each of
the iteration steps minimizes the same functional, independently, and
that this functional is bounded from below as a sum of two non--negative
terms. The only point to notice is that when $p(y|\x)$ is fixed we are
back to the rate distortion case with fixed distortion matrix
$d(x,\x)$. The argument in \cite{CT84} for the BA algorithm applies
here as well. On the other hand we have just shown that the third
equation
minimizes the expected relative entropy without affecting the mutual
information $I(\X;\tX)$. This proves the convergence of the
alternating iterations. However, the situation here is similar to the
EM algorithm and the functional $\F\left[p(\x|x);p(\x);p(y|\x)\right]$
is convex in each of the distribution independently but {\em not} in
the product space of these distributions. Thus our convergence proof does
not imply uniqueness of the solution.
\end{proof}

\subsection{The structure of the solutions}

The formal solution of the self consistent equations, described above,
still requires a specification of the structure and cardinality of
$\tX$, as in rate distortion theory.
For every value of the Lagrange multiplier $\beta$ there are
corresponding values of the mutual information $I_X \equiv I(X,\tX)$,
and $I_Y \equiv I(\tX,Y)$  for every choice of the cardinality of $\tX$.
The variational principle implies that
\beq{anneal1}
\frac{\delta I(\tX,Y)}{\delta I(X,\tX)} = \b1 > 0 ~,
\eeq
which suggests a {\em deterministic annealing} approach.
By increasing the value  of $\beta$ one can move
along {\em convex} curves in the ``information plane'' $(I_X, I_Y)$.
These curves, analogous to the rate distortion curves,
exists for every choice of the cardinality of $\tX$. The solutions of
the self consistent equations thus correspond to a family of such
annealing curves, all starting from the (trivial) point $(0,0)$ in the
information plane with infinite slope and parameterized by  $\beta$.
Interestingly, every two curves in this family separate (bifurcate) at
some finite (critical) $\beta$ through a second order
phase transition. These transitions form a hierarchy of
relevant quantizations for different cardinalities of $\tX$,
as described in \cite{Bialek_Tishby99,SloTish99,PTL93}.

\subsection*{Further work}
The most fascinating aspect of the information bottleneck principle
is that it provides a unified framework for different information
processing
problems, including prediction, filtering and learning
\cite{Bialek_Tishby99}.
There are already several successful applications of
this method to various ``real'' problems, such as semantic clustering
of English words \cite{PTL93}, document classification \cite{SloTish99}, neural
coding, and spectral analysis.

\subsubsection*{Acknowledgements}
Helpful discussions and insights on rate distortion theory with
Joachim Buhmann and Shai Fine are greatly appreciated.  Our collaboration
was facilitated in part  by a grant from the US--Israel Binational
Science Foundation (BSF).

\end{document}